\documentclass[prl,aps,floats,twocolumn,showpacs,nofootinbib]{revtex4}

\usepackage{amssymb}
\usepackage{amsmath}

\usepackage{graphicx}
\usepackage{graphics}
\usepackage{dcolumn}
\usepackage{color}
\usepackage{rotate}
\usepackage{fancyhdr} 
\usepackage{hyperref}
\usepackage{indentfirst}
\newcommand{\ra}[1]{\renewcommand{\arraystretch}{#1}}

\usepackage{umoline}
\usepackage{ulem}

\def\be{\begin{equation}}
\def\ee{\end{equation}}

\begin{document}

\title{A Strategy to Minimize Dust Foregrounds in B-mode Searches}
%\title{Searching for B-modes: The Importance of Being Cleanest}

\author{Ely D.\ Kovetz and Marc Kamionkowski}

\affiliation{Department of Physics and Astronomy, Johns
     Hopkins University, 3400 N.\ Charles Street, Baltimore,
     Maryland 21218, USA} 

\date{\today}

\begin{abstract}
The Planck satellite has identified several patches of sky with
low polarized dust emission, obvious targets for searches for
the cosmic-microwave-background (CMB) B-mode signal from
inflationary gravitational waves.  Still, given the Planck
measurement uncertainties, the polarized dust foregrounds in
these different candidate patches may differ by an order of
magnitude or more.  Here we show that a brief initial experiment
to map these candidate patches more deeply at a single high
frequency can efficiently zero in on the
cleanest patch(es) and thus improve significantly the
sensitivity of subsequent B-mode searches.  A ground-based
experiment with current detector technology operating at
$\gtrsim220\,{\rm GHz}$ for 3 months can efficiently identify a
low-dust-amplitude patch and thus improve by up to a factor $2$ or $3$
on the sensitivity to cosmic B modes of the subsequent
lower-frequency deep integration.  A balloon experiment with
current detector sensitivities covering the set of patches and
operating at $\sim\!350\,{\rm GHz}$ can reach a similar result
in less than two weeks.  This strategy may prove crucial in accessing 
the smallest gravitational-wave signals possible in large-field
inflation.  The high-frequency data from this exploratory
experiment should also provide valuable foreground templates to
subsequent experiments that integrate on any of the candidate
patches explored.
\end{abstract}

\maketitle

A significant effort is now underway to detect the B-mode signal
\cite{Kamionkowski:1996zd} induced by inflationary gravitational
waves \cite{Abbott:1984fp} in the polarization of the cosmic
microwave background (CMB).  Any such search must, however,
distinguish the cosmic signal from the contamination by
Galactic foregrounds, in particular polarized emission from 
dust, as recent experience has shown
\cite{Ade:2014xna,Flauger:2014qra,Mortonson:2014bja,BICEPPlanck}.
Recent data from the Planck satellite on
polarized dust emission at 353 GHz has shown that dust
polarization may be higher than had been expected, even in
low-intensity regions at high Galactic latitudes
\cite{PlanckSep2014}.  The data also indicate that the
amplitude of dust-intensity fluctuations in a given patch of sky
is not necessarily a reliable proxy for the dust-polarization
fluctuations, as had been assumed before.

The importance of the detection of inflationary B modes for
cosmology motivates us to reconsider, in the light of this
new information, every possible avenue to reduce the 
contamination from dust in CMB B-mode maps and
thus maximize the sensitivity of B-mode experiments.  An
unambiguous detection will of course require multifrequency
component separation, and a putative cosmic signal must have the
correct statistical properties \cite{Kamionkowski:2014wza}.
Still, several experiments that focus on a small ($\sim400$
square degrees) patch of sky \cite{Jaffe:2000yt} (e.g., BICEP3,
PolarBear.....), to target the recombination peak in the B-mode
power spectrum, will do better if they integrate on cleaner
(i.e., lower dust-polarization foreground) patches of sky.  An
obvious question for these experiments is which of the $\sim100$
$\sim$400-square-degree patches of the sky to survey.  In the
past, this selection was based on available information on
dust-intensity fluctuations.  Now that we have 353-GHz
polarization maps from Planck \cite{PlanckSep2014}, the
selection will be based largely upon that information.

The silver lining for B-mode searches of the Planck
dust-polarization maps is the identification of a
handful of patches that may potentially have much
lower dust-polarization amplitudes than the average patch. The
crucial point, though, is that due to the limited Planck
sensitivity, the measurements in these low-dust-polarization
patches are noise-dominated, and it is impossible to determine
at this point which of these candidate patches is cleanest of
dust-polarization foregrounds.  The dust amplitudes could
plausibly vary by an order of magnitude or even more.  

Small-sky experiments are hence faced with a difficult
choice: which of Planck's best patches should be chosen
for the deep integration required to optimize the sensitivity to
B modes?  Should they choose one at random and hope for the
best? or perhaps split the total observation between multiple
patches?

An alternative strategy, which we consider here, is to conduct a
brief initial exploratory survey, at a high (dust-dominated)
frequency, of the few relatively clean Planck patches
to identify the cleanest of them.  As we will show, a B-mode
experiment that then focuses on this cleanest patch may
potentially be $\sim20\%-66\%$ more sensitive than it would be
without this initial stage of exploration.  More precisely,
we use as a figure of merit the lowest upper bound
achievable by a null experiment on the tensor-to-scalar ratio $r$.
We then compare the bound obtained in several different scenarios for
the exploration phase to that obtained with no exploration.
We show that the $\sim20\%-66\%$ improvement can be achieved
with an initial 3-month ground-based survey at 220 GHz, with current
detector technology, or with a two-week balloon experiment at
353 GHz.  We also consider briefly several strategies to
optimize the initial stage of exploration.  We will see that the
proposed initial exploration may allow experiments to more
efficiently access the lowest gravitational-wave amplitudes expected in
large-field models \cite{Efstathiou:2005tq,Lyth:1996im}.

We use standard techniques (see, e.g., Ref.~\cite{Wu:2014hta})
to estimate the sensitivity of a given experiment to B modes in
the presence of foregrounds and lensing.  The Fisher forecast
for the error in the measurement of the amplitude $A$ of a power
spectrum $C_\ell$, with some assumed multipole-moment $\ell$
dependence, is \cite{Jaffe:2000yt,Wu:2014hta}, 
\be\label{eq:smallamp}
     \frac{1}{\sigma_A^2} = \sum_\ell \left( \frac{ \partial
     C_\ell }{ \partial A} \right)^2 \frac{1}{\sigma_\ell^2}.
\ee
We assume that the likelihood function
is Gaussian in the vicinity of its maximum
\cite{Jungman:1995bz,Zhao:2008re}. To emphasize the rough nature
of this approximation we use quotation marks around
``1$\sigma$'' when referring to the error $\sigma_A$ in the
amplitude $A$.

We now consider an experiment to identify through measurements
at a single high (dust-dominated) frequency $\nu_{\rm dust}$ the
cleanest of several candidate patches of sky.  We assume that 
the observation time is divided into several steps, each
dedicated to a different patch of sky.  We then need to estimate
the {\it dust amplitude} $A$ in a targeted patch within the
observation time allotted to a {\it single step} of the
exploration stage.  For a given sky coverage $f_{\rm sky}$, the
``1$\sigma$'' error to the value of $A$ estimated from an individual 
multipole moment $\ell$ is
\cite{Jaffe:2000yt,Knox:2002pe,Kesden:2002ku},
\be
  \sigma_\ell^{\widehat{A}} = \sqrt{\frac{2}{ f_{\rm
  sky}(2\ell+1)}}\left(\alpha C^L_\ell +
  f_{\rm sky} w^{-1}(t_{\rm step}) e^{\ell^2\sigma_b^2}\right),
\ee
where $\widehat{A}$ denotes the estimated value of $A$, 
$C^L_\ell$ is the lensing B-mode contribution and $w^{-1}(t_{\rm step})$
is the inverse weight per solid angle given an observation time
$t_{\rm step}$.  The quantity $1-\alpha$ parametrizes the level
of de-lensing \cite{Kesden:2002ku,Sigurdson:2005cp,Smith:2010gu} that is
applied to the data. To be conservative, we assume
$\alpha=1$ (no de-lensing) unless stated otherwise.  The total
``$1\sigma$'' error in the measurement of $\widehat{A}$ over a
time $t_{\rm step}$ is thus,
\be
     \sigma^{\widehat{A}}=\left[\frac{f_{\rm sky}}{2}
     \sum\limits_{\ell_{\rm min}}^{\ell_{\rm max}}
     \frac{(2\ell+1)(\tilde{C}^D_\ell)^2}{\left(
     C^L_\ell+f_{\rm sky}w^{-1}(t_{\rm step})e^{\ell^2
     \sigma_b^2}\right)^2}\right]^{-\frac{1}{2}},
\label{eq:estimatorerror}
\ee
where $\tilde{C}_{\ell}^D=C^D_\ell/A = 2\pi\ell^{-m}/[\ell(\ell+1)]$ 
encodes the $\ell$ dependence of the dust power spectrum (which we 
assume to be the same across the sky, as suggested by
Ref.~\cite{PlanckSep2014}) and $\ell_{\rm min}=40$ is the lowest
multipole included in the analysis. We emphasize that we are interested 
in the dust amplitude $A$ in the {\it particular patch} we are considering for 
prolonged observation, and hence no cosmic-variance term appears in 
Eq.~(\ref{eq:estimatorerror}).

After the optimal patch is identified through exploration, a
prolonged stage of integration is performed over the chosen
patch at a CMB-domimated frequency $\nu_{\rm CMB}$ (and as we
discuss below, perhaps additional frequencies).  To estimate the
gain achieved by exploration, we compare
the sensitivity to primordial B modes from the integration.
The smallest amplitude in a patch $p$ detectable at ``1$\sigma$'' 
(with a total observation time $T$ spent on the chosen patch),
according to Eq.~(\ref{eq:smallamp}), is then \cite{Wu:2014hta}
\be
\resizebox{\hsize}{!}{$
     \sigma_p^r=\left[\frac{f_{\rm sky}}{2}
     \sum\limits_{\ell_{\rm min}}^{\ell_{\rm
     max}}\left(\frac{\sqrt{(2\ell+1)}\tilde{C}^B_\ell}{\alpha
     C^L_\ell+\beta A_p(\nu_{\rm CMB})\tilde{C}^D_\ell+f_{\rm
     sky}w(T)^{-1}e^{\ell^2\sigma_b^2}}\right)^2\right]^{-\frac{1}{2}}$}.
\label{eqn:sigmaprone}
\ee
where $A_p(\nu_{\rm CMB})$ is the dust amplitude in the patch $p$ 
(extrapolated to frequency $\nu_{\rm CMB}$),
$\tilde{C}^B_\ell$ denotes the $\ell$ dependence of the IGW B-mode
power spectrum and $\beta$ is the fraction of the foreground
left in the $\nu_{\rm CMB}$ map, after component separation is
performed (in a multifrequency experiment); $\beta=1$ with only
one $\nu_{\rm CMB}$.  We set the sample variance of 
the primordial signal to zero in this expression (to compare
with the null hypothesis).

An important issue if that of the bias stemming from dust power 
in cases where it is not cleaned (e.g. in a single-frequency experiment).
One could attempt to resolve this in a number of ways. First, as discussed in
\cite{bandits}, it is possible to fit simultaneously for $r$ and $A$ on the {\it integration} data, 
using the different $\ell$-dependence of dust and gravitational waves. Another
way is to employ the {\it exploration} measurements on the same patch of sky, 
either for standard component separation, or as a means of inferring the dust amplitude
which could then be extrapolated to the {\it integration} frequencies and used to 
correct the data. Naturally, full component separation with multi-frequency data is
the superior method, and we will assume that is possible for most of the experimental
setups we consider below.

The instrumental noise in a CMB-polarization experiment is determined 
by the detector-array sensitivity $s=s_{\rm det}/\sqrt{N_{\rm det}}$ (where $s_{\rm det}$ is the 
noise-equivalent temperature NET of each detector and $N_{\rm det}$ is the number of 
detectors in the array), the angular resolution $\theta_{\rm fwhm}$, 
the sky coverage $f_{\rm sky}$, and the total observation time $T$ 
(which is reduced in practice by the observing efficiency). 
The pixel noise $\sigma_{\rm pix}$ is then determined by
$\sigma_{\rm pix}=s/\sqrt{t_{\rm pix}}$, where $t_{\rm
pix}=T/N_{\rm pix}$ is the observation time dedicated to each
pixel. Defining the inverse weight $w^{-1}(T)=4\pi
s^2/T$ per solid angle, the angular power spectrum of the
instrumental noise, assuming the experimental beam is
approximately Gaussian in shape, is given by
\cite{Tegmark:1997vs}
\be
C^N_{\ell}=\frac{\Omega\sigma_{\rm pix}^2}{N_{\rm pix}}e^{\ell^2\sigma_b^2} = 
\frac{\Omega s^2}{T}e^{\ell^2\sigma_b^2}=f_{\rm sky}w^{-1}(T)e^{\ell^2\sigma_b^2},
\ee
where $\Omega=4\pi f_{\rm sky}$ and $\sigma^2_b=\theta^2_{\rm fwhm}/(8\ln{2})$.

In Table \ref{ExpTable} we list the parameters of the
instruments we consider for the exploration experiment and the
subsequent B-mode integration, where for all experiments we fix
the NET per detector to be $s_{\rm det}=480\,{\rm \mu K \sqrt{s}}$
(which is already surpassed by most current generation
instruments).  We note that these parameters are similar to
those considered, e.g., in Ref.~\cite{Wu:2014hta}. Since dust is
bright at high frequencies, current-generation ground-based
or balloon experiments with $N_{\rm det}=O(10^3)$ will most
likely suffice.  
For the prolonged integrations we also consider more sensitive
experiments (see Ref.~\cite{Wu:2014hta}), with $N_{\rm det}$ as
high as $10^5$. We set the minimum angular resolution to $\theta_{\rm
fwhm}=30'$ for all experiments, so that the primordial peak at
$\ell\sim80$ may be resolved.Optimistic de-lensing
prospects such as we consider below will require higher resolution E-mode 
polarization data along with decent lensing potential estimates (e.g.\ via tracers such as the CIB) \cite{Smith,Sherwin}.
Finally, we assume an observing efficiency of $20\%$ and a sky 
coverage of $f_{\rm sky}=1\%$.

\begin{center}
\begin{table}[htbp]\centering \footnotesize
\ra{1.3}
\begin{tabular}
{@{}r|p{0.1cm}p{1.4cm}p{1cm}p{2cm}p{1.4cm}@{}}\hline
 && Type & $N_{\rm det}$ & $\nu_{\rm obs}$ [GHz] & $t_{\rm obs}$ [days]  \\ \hline 
Exploration1$\,$ &&Ground  & $10^3$ & 220 & 90   \\
Exploration2$\,$ &&Balloon  & $500$ & 353 & 12 \\
\hline 
Integration1&&Ground  & $10^3$ & 150 & 1000   \\
Integration2&&Ground  & $10^4$ & \{90, 150, 220\} & 1000   \\
Integration3&&Ground  & $10^5$ & \{90, 150, 220\} & 1000  \\
\hline
\end{tabular}
\caption{Table of experimental parameters for the exploration
and integration experiments under consideration.}
\label{ExpTable}
\end{table}
\end{center}

The Planck experiment reported in \cite{PlanckSep2014} the results
of an analysis of its $353\,{\rm GHz}$ maps
at high Galactic latitudes ($|b|>35^{\circ}$), where $352$
patches of $400\,{\rm deg}^2$ were analyzed in the multipole
range $40 < \ell < 370$, to determine the amplitude of dust
polarization. This analysis identified a group of
candidate patches with low dust amplitudes, although these
amplitudes had large uncertainties. 

In Table \ref{DustAmpTable}, we list the amplitudes of Planck's
six best candidates, normalized to the primordial
gravitational-wave B-mode power spectrum at $\ell=80$ and
extrapolated down from $353\,{\rm GHz}$ to $150\,{\rm GHz}$,
assuming a modified blackbody spectrum with spectral index
$\beta_d=1.59$ and $T_d=19.6K$, applying the unit conversion
factors and color correction coefficients as explained in
\cite{PlanckSED}. 

\begin{center}
\begin{table}[htbp]\centering \footnotesize
\ra{1.3}
\begin{tabular}
{@{}r|p{0.1cm}p{1cm}p{0.875cm}p{1.125cm}p{1cm}p{0.875cm}p{1cm}@{}}\hline 
$r_d$ && $0.053$ & $0.027$ & $-0.062$ & $-0.020$ & $0.057$ & $-0.031$\\
$\sigma_d$&& $0.096$ & $0.098$ &$~\,$ $0.052\,$ &  $~\,$ $0.127$ & $0.122$ & $~\,$ $0.121$   \\
\hline
\end{tabular}
\caption{The best-fit dust amplitude $r_d$ and its uncertainty $\sigma_d$ for the six best patches identified by Planck at $353\,{\rm GHz}$. The values shown here are normalized to the amplitude of primordial B-modes at $\ell=80$ and extrapolated to $150\,{\rm GHz}$ assuming a modified blackbody spectrum \cite{PlanckSED}.}
\label{DustAmpTable}
\end{table}
\end{center}

Given the large uncertainties in the measurements of the dust
amplitudes in these six patches, we cannot claim to know how
much better one can do in terms of B-mode detection by locating the
cleanest patch of the sky.  We can, however, study the potential
improvements obtained from locating the cleanest patch of sky by
simulating skies consistent with Planck measurements.

To do so, we generate an ensemble of simulated skies where the
dust amplitudes in the candidate set are drawn from normal
distributions with the best-fit mean and variance reported by
Planck $\mathcal{N}(r_d,\sigma_{r_d})$ (for each patch in each
simulation, we repeat the draw until a positive amplitude is
drawn), and calculate ensemble averages for all desired
quantities. We then use the conversion convention described
above to transform between $150\,{\rm GHz}$ and either
$220\,{\rm GHz}$ or $353\,{\rm GHz}$.

In Table ~\ref{TSratioProspects}, we calculate the lowest
detectable tensor-to-scalar ratio for the worst, mean, and best
foreground amplitude, averaged over the ensemble of simulations,
assuming a single-frequency experiment (where $\beta=1$) as well
as multifrequency experiments, where after component separation
a residual of either $\beta=10\%$ or $\beta=1\%$ is left, with
and without de-lensing (we consider either no de-lensing,
$\alpha=1$, or an optimistic $90\%$ de-lensing, 
$\alpha=0.1$). The parameters for these B-mode-integration
experiments are taken from Table \ref{ExpTable}. 

We should mention that there is some redundancy in the scenarios 
covered in Tables \ref{ExpTable} and \ref{TSratioProspects}. With no
de-lensing, for example, the third integration experiment will be lensing-noise limited
in a small patch of sky and will not exceed the sensitivity of the second experiment,
unless a larger area of sky is covered, to reduce the sample-variance error.
For simplicity, we neglect such optimizations here as our main concern is the 
effect of foreground reduction (by means of avoidance) and our figure-of-merit 
is the lowest bound achievable on the tensor-to-scalar ratio under the null hypothesis.
Naturally, these considerations should play a dominant role when planning an actual 
experimental strategy for a given instrument (along with various other considerations, 
e.g.\ constraints on the observing efficiency in different regions of the sky).

Clearly, there is room for improvement if cleaner patches can be
identified before the prolonged integration is done. We
therefore turn to investigate how best to identify the cleanest
patch.

\begin{center}
\begin{table}[htbp]\centering \footnotesize
\ra{1.3}
\begin{tabular}
{@{}r|p{0.1cm}p{0.5cm}|p{0.75cm}p{0.75cm}p{0.8cm}|p{0.975cm}p{0.975cm}p{0.975cm}@{}}\hline
 && $\beta$ &Worst & Mean & Best  &Worst $\alpha=0.1$ & Mean $\alpha=0.1$ & Best  $\alpha=0.1$ \\ \hline 
Integration1&& 1  & 0.034  & 0.022 & 0.011   & 0.03 & 0.018 & 0.009 \\
Integration2&& 1 & 0.025& 0.014 & 0.006 & 0.019 & 0.01 & 0.003   \\
Integration2&& 10\% & 0.007 & 0.005 & 0.004 & 0.0033 & 0.0022 & 0.0011   \\
Integration3&& 1 &0.023 & 0.013  & 0.005 &0.016 & 0.008 & 0.002  \\
Integration3&& 10\% & 0.006 & 0.005 & 0.003 & 0.0024 & 0.0014 & 0.0006   \\
Integration3&& 1\% & 0.0033 & 0.0031 & 0.0029  & 0.0007 & 0.0005  & 0.0004  \\
\hline
\end{tabular}
\caption{Table of expected lower bounds on $r$ with different
integrations, with no exploration.}
\label{TSratioProspects}
\end{table}
\end{center}

Given a set of candidate patches for which we have rough
dust-amplitude estimates, we can choose to forgo exploration
(and just choose the best candidate according to the
available data, uncertain as it may be) or to pursue
a uniform exploration of all patches so as to lower the measurement
uncertainty in the dust amplitude and thus allow a more
substantiated choice.  Alternatively, one could adopt an
adaptive strategy, whereby the allocation of exploration time is
determined on-the-fly according to the gathered results.  In
this case, patches that acquired data indicate are most likely
inferior are discarded early on, leaving more time to explore more deeply
the remaining patches and thus zero in efficiently on the best
ones.

In Ref.~\cite{bandits}, we investigated a variety of
adaptive-survey strategies for ground-based single-frequency
experiments (similar to BICEP2).  We showed
how to construct adaptive survey strategies, based on
machine-learning algorithms, to maximize the sensitivity to
primordial B-modes by spending more time observing
lower-foreground patches.  We concluded that a wisely chosen
adaptive-survey strategy could improve the upper limit to B
modes (assuming a null experiment) by a factor of 2--3.

Here, though, we consider a limited fixed-time exploration at a
dust-dominated high frequency which is then followed by a
prolonged integration at CMB frequencies.  In this case, the
advantage of adaptive strategies is limited, first because the
measurements are easier and secondly because there is no price
paid in terms of integation time (the integration is done
separately, by an independent CMB-frequency
measurement). Nevertheless, we compare the performance of the
adaptive Upper Confidence Bound (UCB) adaptive-survey algorithm
(which was found in Ref.~\cite{bandits} to be quite effective)
to a more naive uniform exploration to study the improvements that
can be made if a given experiment can move quickly between patches
at low cost.

\begin{figure}
\includegraphics[width=0.95\linewidth]{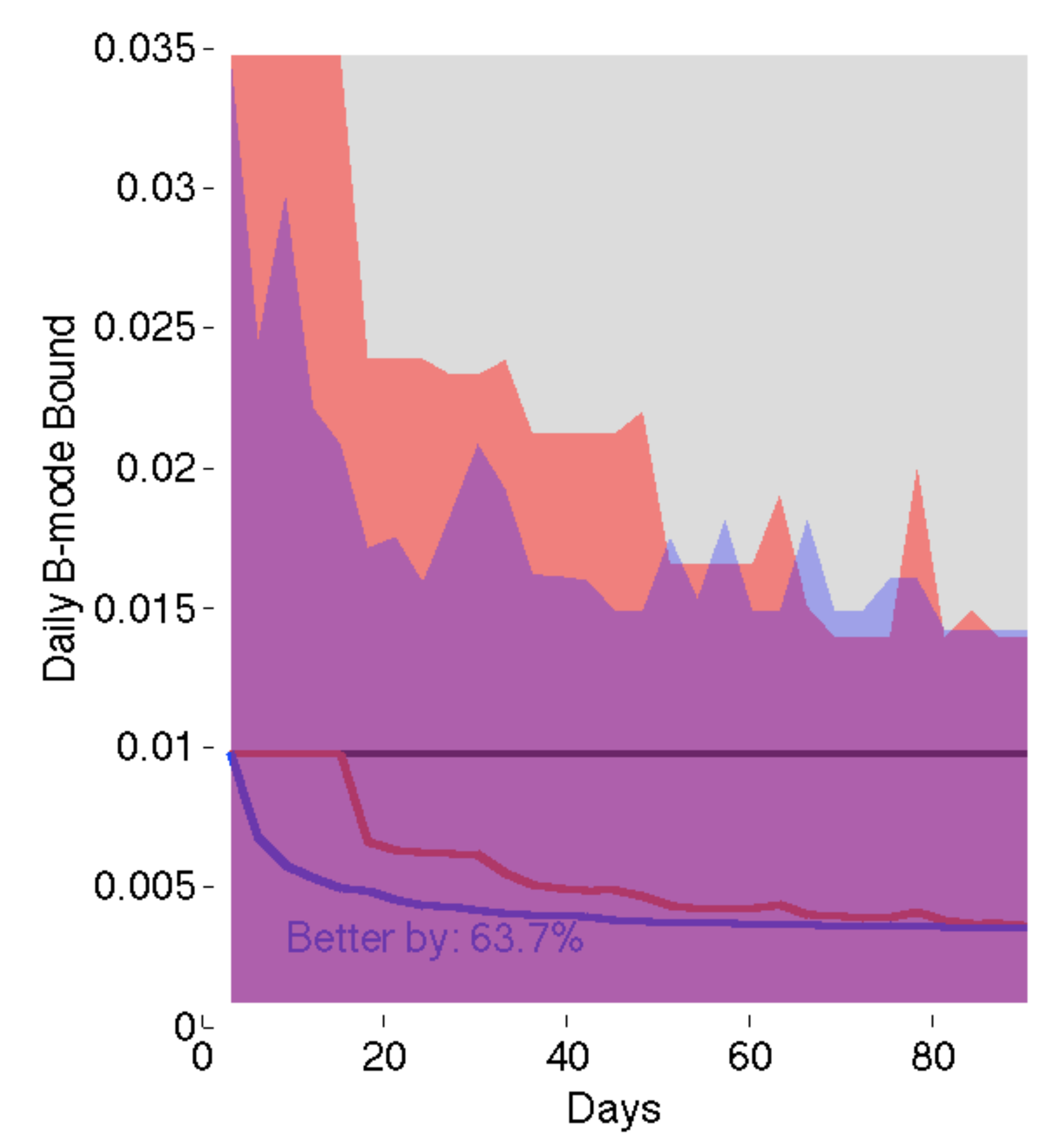}
\includegraphics[width=0.85\linewidth]{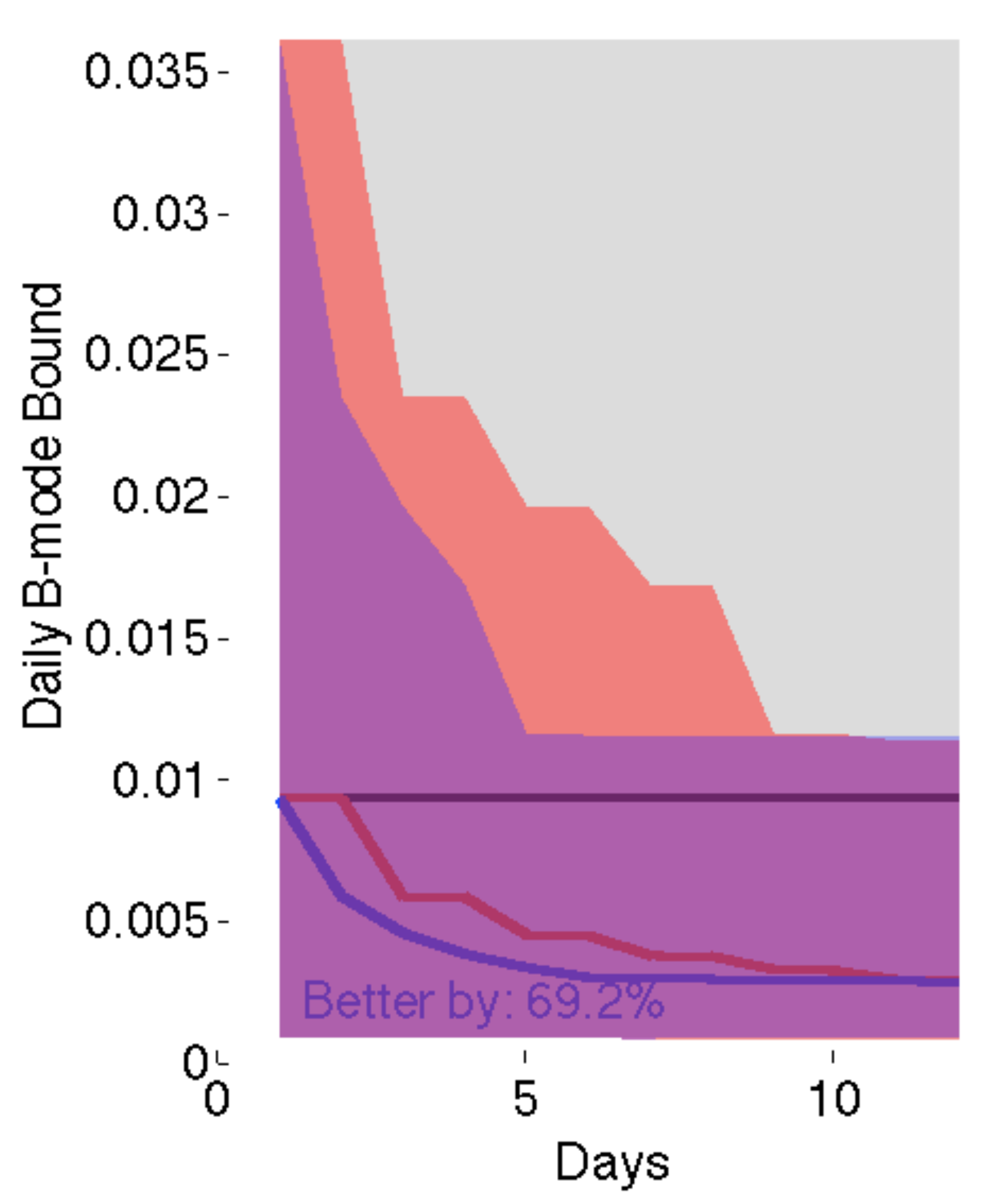}
\caption{The projected ``$1\sigma$" bound on
the tensor-to-scalar ratio $r$ with the instrument  
\textit{Integration2} with $\beta=10\%$ and $\alpha=0.1$ (see Table \ref{ExpTable}), 
achievable with the information obtained from $x$ days of exploration. 
We plot the results given a ground-based 
experiment (\textit{Top}), or a balloon-borne one (\textit{Bottom}), 
both with parameters taken from Table \ref{ExpTable}. Three methods are 
compared: no exploration (solid black line); consecutive equal-time
observation (solid red line); an adaptive method (UCB) (solid
blue line). The ranges of performance are also shown 
(in corresponding color shades) to
demonstrate the worst case for each method. After all patches
are observed for $1/6$th of the total exploration time, the
uniform method reaches the same bound as the adaptive one.} 
\label{fig:Exploration}
\end{figure}

In Fig.~\ref{fig:Exploration}, we track the average $1\sigma$
bound on the tensor-to-scalar ratio $r$ that may be reached by a
subsequent integration experiment (we plot the results for 
the instrument \textit{Integration2} with $\beta=10\%$ and $\alpha=0.1$), given
each additional day of exploration. At each point in time, we plot the achievable
bound on the tensor-to-scalar ratio $r$ assuming it observes the patch 
with the lowest-measured dust amplitude by the exploration experiment up to that stage.
In the top panel we show results from a ground-based experiment and 
in the bottom from a balloon instrument, with parameters from Table \ref{ExpTable}.
As explained above, we compare three strategies. The black solid line shows the average
daily bound assuming a patch is chosen at random and no
exploration performed (hence no improvement). Its range of
performance is shown in grey. The red solid line shows the daily
bound as the patches are consecutively observed, each for a period of $15$
days in the case of the ground-based instrument (or two days for the balloon). 
During the first observation period, there would be no gain compared to
choosing at random, as expected. Then, as each additional patch
is observed, the projected bound on $r$ improves as more information is
gathered to allow an intelligent choice of patch for
integration. The range of performance of this method is shown
in red. Finally, the blue solid line shows the daily bound with
the UCB method, which converges onto cleaner patches much more
quickly, but requires a flexible experiment which can switch
between patches every three days (or a single day, for the
balloon experiment).
As can be seen, the worst case when adopting this strategy is almost 
always better than with the other methods.

How well will the B-mode searches (the
integrations) perform if they are preceded by a stage of exploration as
described above? In Table \ref{TSratioProspectsExploration} we
present the average and worst-case B-mode limits achievable by
the same integration experiments considered in Table
\ref{TSratioProspects}, after the ground-based exploration
experiment in Table \ref{ExpTable} has been 
performed to identify the optimal patch for observation in each
simulation.  (The balloon-borne exploration yields very similar
results, so we do not show them separately). We see that the
improvement compared to the \textit{mean} without exploration
ranges from $20\%$ to $66\%$ in almost all experiments, depending on
the level of foreground subtraction and the amount of
de-lensing.  Another advantage of the exploration stage is
that the worst case is now greatly reduced compared to the
non-exploration case, as can be seen clearly.

The only case in which there is only minor
improvement, $<5\%$, is with $N_{\rm det}=10^5$, $1\%$
foreground residuals, and no de-lensing. There
is no improvement here as both the foreground residuals and
instrumental noise are lower than the lensing B modes for all
six patches and hence there is no gain by focusing on cleaner
ones. Such a tactic, though, cannot reach the value
$r\simeq0.002$ that is targeted to fully probe the parameter
space for large-field-inflation.

\begin{center}
\begin{table}[htbp]\centering \footnotesize
\ra{1.3}
\begin{tabular}
{@{}r|p{0.1cm}p{0.5cm}p{0.8cm}p{1.8cm}|p{1cm}p{2cm}@{}}\hline
 && $\beta$ &Worst & Mean  ($\Delta(\%)$) &Worst $\alpha=0.1$ & Mean $\Delta(\%)$) $\alpha=0.1$  \\ \hline 
Integration1&& 1  & $0.025$ & $0.012$ $(39\%)$ &  $0.0213$ & $0.0098$  $(45\%)$ \\
Integration2&& 1 & $0.019$ & $0.007$ $(51\%)$ &  $0.0142$ & $0.0036$  $(64\%)$ \\
Integration2&& 10\% & $0.006$ & $0.004$ $(21\%)$ & $0.0028$   & $0.0013$ $(39\%)$ \\
Integration3&& 1        & $0.015$   & $0.006$  $(53\%)$ & $0.0099$     & $0.0027$  $(66\%)$ \\
Integration3&& 10\% & $0.005$   & $0.003$  $(23\%)$ & $0.0018$  & $0.0007$  $(50\%)$ \\
Integration3&& 1\%   & $0.003$ & $0.003$  $(<\!5\%)$ &  $0.0007$    & $0.0004$ $(20\%)$ \\
\hline
\end{tabular}
\caption{Table of expected lower bounds on $r$ with different
integration experiments, following a stage of
exploration. Results are shown for three integration
experiments with parameters listed in Table \ref{ExpTable}, 
assuming different levels of foreground residuals
(from $100\%$ to $1\%$) and of de-lensing ($\alpha=1$ for
no-de-lensing or $\alpha=0.1$ for $90\%$ de-lensing). The
improvement achievable on average compared to the
non-exploration case is shown in parentheses.}
\label{TSratioProspectsExploration}
\end{table}
\end{center}

To conclude, we have considered the
benefits for B-mode searches of an initial exploratory experiment
to measure at a single high (dust-dominated) frequency the
dust-polarization amplitudes in several candidate
low-dust-polarization patches identified by Planck.  We have
shown that the sensitivity of the subsequent B-mode searches can
be improved, with the identification of the cleanest patch with
such an exploration, so that the upper bound achievable on the 
tensor-to-scalar can be decreased by as much as $\sim20\%-66\%$
(or as much as a factor of $3$ lower). We showed that
adaptive-survey strategies for the exploration experiment can
allow the cleanest patch to be identified a bit more efficiently
if the experiment has the flexibility to rapidly change
targets.  However, the improvement over a simpler rigid
exploration is not particularly significant.

Although we have assumed in the analysis just one deep
integration experiment, there are likely to be a significant number of
independent projects, over the coming years, seeking inflationary B modes.
The single exploration experiment to identify the cleanest patch is
therefore likely to benefit all subsequent experiments, allowing each
of them $\sim20-66\%$ improvements in sensitivity.  
The deep data from the exploratory experiment (particularly one at 353
GHz) should also provide valuable foreground templates for
subsequent CMB experiments on {\it any} of the candidate patches
explored. Finally, such data may also be useful to improve our 
theoretical understanding of dust-polarization by clarifying some of the 
open questions and issues raised by Planck \cite{PlanckSep2014}.

All of this improvement comes at the expense of just
90 days of observation from the ground at 220 GHz or two weeks
by a balloon operating at 353 GHz. This price seems affordable.

\smallskip

We thank the anonymous referee for important comments which 
helped improve the clarity of our paper.
This work was supported by the John Templeton Foundation, the Simons
Foundation, NSF grant PHY-1214000, and NASA ATP grant NNX15AB18G.

\end{document}